\documentclass[12pt]{revtex4}
\usepackage{amssymb}
\usepackage{graphicx}
\usepackage{epstopdf}

\def\be{\begin{equation}}
\def\eec{, \end{equation}}

\def\be{\begin{equation}}
\def\ee{ \end{equation}}

\def\Msun{M_{\odot}}

\begin{document}
\title{Mass of highly magnetized white dwarfs exceeding the Chandrasekhar limit: An analytical view}
\smallskip\smallskip 
\author{Aritra Kundu}
\affiliation {Department of Physics, Indian Institute of Technology, Kanpur}
\email {aritrakundu@gmail.com, aritrak@iitk.ac.in}
\author{Banibrata Mukhopadhyay}
\affiliation{Department of Physics,  Indian Institute of
Science,  Bangalore 560012,  India}
\email{bm@physics.iisc.ernet.in}  

\vskip10cm 
\begin{abstract}
In recent years a number of white dwarfs has been observed with very high surface magnetic fields. We can 
expect that the magnetic field in the core of these stars would be much higher ( $\sim 10^{14}~G$). 
In this paper, we analytically study the effect of high magnetic field on relativistic cold electron, and 
hence its effect on the stability and the mass-radius relation of a magnetic white dwarf. 
In strong magnetic fields, the equation of state of the Fermi gas is modified and Landau quantization 
comes into play. For relatively very high magnetic fields (with respect to the energy density of matter) 
the number of Landau levels 
is restricted to one or two. We analyse the equation of states for magnetized electron degenerate gas
analytically and attempt to understand the conditions in which transitions from the zero-th Landau level to first Landau level occur. We also find the effect of the strong magnetic field on the star collapsing to a white dwarf, 
and the mass-radius relation of the resulting star. We obtain an interesting theoretical result that 
it is possible to have white dwarfs with mass more than the mass set by Chandrasekhar limit. 
\end{abstract}

\maketitle

{\bf Keywords:} white dwarfs, degenerate Fermi gases, equations of state of gases,
Landau levels, stellar magnetic fields  \\

{\bf PACS No. :} 97.20.Rp, 67.85.Lm, 51.30.+i, 71.70.Di, 97.10.Ld \\

\newpage
\section{Introduction}\label{intro}
The origin of high magnetic fields in compact stars is explained by the fossil field hypothesis, as 
proposed by Ginzburg \cite{ginz} and Woltjer \cite{wolt}. The magnetic flux $\phi_b \sim 4\pi BR^2 $ of a 
star is conserved during its evolution, thus a degenerate collapsed star is expected to have a very high
magnetic field if the original star had a magnetic field, say, $B \sim 10^8 ~G $ \cite{shapiro} \cite{spitzer}.  
Neutron stars were of the main interest in this category, but in recent years some white dwarfs have been 
found to have quite high surface magnetic fields with field strength varying from 
$10^6 ~G$ to $10^9~ G $ \cite{mwd1,mwd2,mwd3,mwd4}. 
It is highly intuitive that the fields near the core of the white dwarf would be much higher. 
Ostriker and Hartwick \cite{osha} constructed models of white dwarf with central magnetic field $10^{12}~ G$ 
but a much smaller field in surface. The maximum limit of the field strength is set by the virial theorem 
\be 2T+W+3U+{\cal M}=0 \eec  where $T$ is the total kinetic energy, $W$ the gravitational potential energy, 
$U$ the total internal energy and $\cal M$ the magnetic energy. Since $T$ and $U$ are both positive definite,
the maximum magnetic energy is always less than the total gravitational energy in equilibrium. 
For a star of mass $M$ and radius $R$ this gives \be B_{max} \sim 2\times 10^8\left(\frac{M}{M_\odot}\right)
\left(\frac{R}{R_\odot}\right)^{-2} ~G,\ee where $M_\odot$ and $R_\odot$ respectively denote solar mass and 
radius. For a white dwarf this sets the limit of  $B \sim 10^{12} ~G  $ at the center but relatively 
lower fields outside. 

The mass-radius relation for a non-magnetic relativistic white dwarf as determined by Chandrasekhar 
\cite{chandra} sets its maximum mass to $1.44~\Msun$ such that the electron degeneracy pressure is adequate for 
counterbalancing the gravitational collapse. This mass limit is strengthened by higher central densities 
\cite{shapiro}. The effect of \textit{weak} magnetic field $B\leq10^{13}~G$ on white dwarfs was studied 
by Suh et al. \cite{suh} by applying Euler-MacLaurin expansion on the equation of state for a degenerate 
electron gas in a magnetic field \cite{slai}. They found that both the mass and radius of the white dwarf 
increase in presence of magnetic fields. 

In this report we analyse the case of relativistic white dwarfs with magnetic fields $B~>~10^{13}~G$. 
In order to have a white dwarf of such magnetic field, the original solar type star should have a magnetic 
field $\sim 10^9~G$, from the flux freezing theorem. Existence of such stars is not ruled out \cite{shapiro}.
First we analyse the effect of such strong magnetic field on the equation of state of degenerate electron 
gas at zero temperature and analyse some of its properties, subsequently we consider considerable higher 
magnetic field for the same and then study explicitly the mass-radius relation for the case of just one 
Landau level (produced by high magnetic field). The results of higher Landau level occupancy (hence lower 
magnetic field) are addressed quantitatively  from the equation of state. In doing so, we find for central 
densities $\rho \sim 10^{10} ~g/cc$ the maximum mass of the white dwarf can be much greater than the 
Chandrasekhar limit $M \sim 1.44\Msun$. 

\section{ equation of state in high magnetic field}\label{SMFEOS}
\subsection{Density of states}
The energy eigenstates of the free electrons in the magnetic field are quantized to what is known as 
Landau orbitals. The electrons motion perpendicular to the magnetic field is no longer independent but is quantized. This can be seen by solving the time-independent Schr\"odinger
equation of a particle in a magnetic field $B$ directed along $z-$axis given by \cite{landau}
\be
\hat{H}\Psi=\left[\frac{(p_{x}+eBy/c)^{2}}{2m_e}+\frac{p_{y}^{2}}{2m_e}+\frac{p_{z}^{2}}{2m_e}\right]\Psi-\frac{\mu\sigma}{s}B\Psi=E\Psi, 
\ee
where $p_x,p_y,p_z$ are the components of linear momentum, $y$ is the arbitrary position on the $y-$axis,
$m_e$ and $e$ are the mass and charge of the electron
respectively, $c$ the speed of light, $\mu$
the magnetic moment, $s$ the magnitude of spin, $\sigma =\pm \frac{1}{2}$, $E$ the energy eigenvalue.
We obtain the classical quantization of energy
levels of the system given by  
\be
E_{\nu}=\nu\hbar\omega_{H}+\frac{p_{z}^{2}}{2m_e} 
\eec
where $\omega_{H}=\frac{eB}{m_ec}$, is the critical cyclotron frequencies at which quantization occurs,
$\hbar$ the Planck's constant,
and $\nu=(l+\frac{1}{2}+\sigma)$, gives the Landau levels where $l$ is the principal quantum number for 
the electron.  Relativistically we 
obtain energy by solving Dirac equation in a magnetic field, 
hence the energy above is modified to
\be\label{ree}
E_{\nu}=\sqrt{m_e^{2}c^{4}+p_{z}^{2}c^{2}+2\nu e\hbar Bc}.
\ee
Thus electrons with up ($\uparrow$) spin and down ($\downarrow$) spin have different
energies. We see that the ground state ($\nu=0$) has degeneracy $1$, while the Landau
levels from $\nu=1$ have degeneracy $2$. The spin $\downarrow$ electrons can occupy
Landau levels $\nu=0, 1, 2, 3...$, while the spin $\uparrow$ ones take Landau levels
$\nu=1, 2, 3...$ \cite{bani}.
Note that the motion of the electrons in the
$x-y$ plane is coupled.
Therefore the density of states of the electron
will change, as the motion is restricted and quantized in the plane.
As the motion in the $x-y$ plane is quantized, the phase space occupied
by the electrons will change. The number of states per unit volume in the interval $\triangle p_z$ 
for a given Landau level $\nu$ is  $g_\nu\left(\frac{eB}{h^2 c}\right)\triangle p_z$. 
The modified density of state is then
\be
2\frac{dp_{x}dp_{y}dp_{z}}{h^{3}}\Rightarrow2\frac{eB}{h^{2}c}{\sum_\nu}g_{\nu 0}dp_{z}(\nu)
\eec
where $g_{\nu 0}=(2-\delta_{0,\nu})$, is the degeneracy in each Landau level, and $dp_{z}(\nu)$ is the 
small element of component of momentum in the $z-$direction in the $\nu-$th Landau level. Thus each 
Landau level has its own distribution of states. The
separation between two Landau levels depends on the strength of the
magnetic field. For high magnetic fields the separation of the
Landau levels is large, hence electrons with low energy (non-relativistic)
can only occupy the ground state. As the strength of the magnetic
field decreases, the separation between the levels decreases. Hence
it becomes energetically favourable for the electrons to jump to higher levels, thus the number of occupied
Landau level increases. Similarly in the case of relativistic electrons, if
the magnetic field is low, the separation of the Landau levels is comparable
to the rest mass energy of the electrons, and hence the electrons can freely
move between the Landau levels, thus it will behave as continuum.  But
again in presence of very high magnetic fields, the electrons, although relativistic,
cannot jump to higher levels. The electrons can become relativistic in two different cases:
(1) when the density is high enough such that the mean Fermi energy of the electron exceeds rest 
mass energy of the electron,
(2) when the term in the energy associated with the cyclotron frequency of the electron exceeds the 
rest mass of the electron. 


\subsection{Equation of state}
We now analyse the equation of state for the relativistic electrons in magnetic field, as was obtained 
earlier \cite{slai}.  We
define the Fermi momentum, analogous to that in the non-magnetic case, as 
$p_{F}^{2}=p_{z}^{2}+\frac{2 \nu eB\hbar}{c}$. Therefore
for $\nu$-th Landau level the $z-$component of the momentum is given
by $p_{z}(\nu)=\sqrt{p_{F}^{2}-\frac{2\nu eB\hbar}{c}}$. We introduce a convenient dimensionless parameter 
$x_{F}=\frac{p_{F}}{m_e c}$, as relativity parameter,  and
$B_{c}=\frac{m_e^{2}c^{3}}{e\hbar}=4.14\times 10^{13}~G$, as the critical magnetic field giving rise
to the significant effect due to the Landau quantization. We
further define $\gamma=\frac{B}{B_{c}}$,  and $\lambda=\frac{\hbar}{m_ec}$, the Compton wavelength of the 
electron. Using these definitions we can
write $p_{z}(\nu)$ as
\be
\frac{p_z}{m_ec}=x_{z}(\nu)=x(\nu)=\sqrt{x_{F}^{2}-2\nu\gamma}
\eec
where $x(\nu)$ denotes the relativity parameter of the $z-$component of momentum
for $\nu$-th Landau level. The number density is then given by
\be
n_{e}=\frac{2\gamma }{(2\pi)^{2}\lambda^{3}}\sum_{\nu=0}^{\nu_{max}}g_{\nu 0}\intop dx(\nu)=\frac{2\gamma }{(2\pi)^{2}\lambda^{3}}\sum_{\nu=0}^{\nu_{max}}g_{\nu 0}x(\nu) \label{nde}
\eec
where $\nu_{max}$ is the maximum number
of Landau level that can be filled which is determined by the condition that
$x_{z}(\nu)$ is real. Hence
\be 
\nu\leq  \nu_{max}= {\rm Integer}(\frac{x_{F}^{2}}{2\gamma})= {\rm Integer}(\frac{\epsilon_F^2-1}{2\gamma})
\label{fille}
\eec
where $\epsilon_F=\frac{E_\nu}{m_ec^{2}}$ which is the dimensionless chemical potential 
and from equation (\ref{ree}) $\epsilon_F^{2}=1+x_F^{2}$. 
This is related to matter density by \be\rho=\mu_em_nn_e=n_e\Theta  \label{densitye}\eec where $\mu_e$ is the mean 
molecular weight given by $A/Z$, $Z$ the atomic number, $A$ the mass number, and 
$m_n$ the mass of the neutron. Note that $\Theta =\mu_em_n$ which has the dimension of mass and denoting the effective mass 
per  electron. Its value depends on the constituents of the white dwarf. In our case value of $\mu_e$ is taken to be $2$, 
and hence $\Theta=2m_n$.
The electron energy density at zero temperature is given by \cite{slai}
\be
 E=\frac{2\gamma }{(2\pi)^{2}\lambda^{3}}\sum_{\nu=0}^{\nu_{max}}g_{\nu 0}\intop^{x_F}_0 E_\nu dx(\nu)
\eec
and hence
\begin{equation}
E=\sum_{\nu=0}^{\nu_{max}}\frac{\gamma g_{\nu 0}m_ec^{2}}{(2\pi)^{2}\lambda^{3}}\left[x(\nu)\epsilon_{F}+(1+2\nu\gamma)\ln[\frac{x(\nu)+\epsilon_{F}}{\sqrt{1+2\nu\gamma}}]\right].\label{energye}
\end{equation}
Then the pressure
of the Fermi gas at zero temperature can be found from the relation
\be
P=\sum\epsilon_{F}x(\nu)-E
\eec
which gives the relation
\be
P=\sum_{\nu=0}^{\nu_{max}}\frac{\gamma g_{\nu 0}m_ec^{2}}{(2\pi)^{2}\lambda^{3}}\left[x(\nu)\epsilon_{F}-(1+2\nu\gamma)\ln[\frac{x(\nu)+\epsilon_{F}}{\sqrt{1+2\nu\gamma}}]\right]\label{eose}.
\ee

\section{one level and the two level systems and equation of states}

First we consider the one level system in which only the ground level is occupied. 
This is possible at a very high magnetic field.
Expanding equation (\ref{nde}) to first term (for $\nu=0$) and with equation (\ref{densitye}) we obtain
\be \rho=\frac{\gamma \mu_e m_n}{2\pi^2\lambda^3}x(0)=\frac{\gamma \mu_e m_n}{2\pi^2\lambda^3}x_F\eec
from which we further obtain 
\be 
x_F=\frac{2\pi^2\lambda^3}{\gamma\mu_e m_n}\rho=~\frac{\rho}{K}
\eec
where $K=\mu_e m_n \frac{2 \gamma}{(2\pi)^2\lambda^3}$. Here we obtain $x_F$ in terms of density and 
hence the relation of Fermi energy to the density of state. As we increase the density of the matter, we in 
turn increase the Fermi energy associated with it, hence electrons acquire probability to
jump to higher Landau levels and 
therefore $\nu$ can not be set $0$. 
From equation (\ref{fille}) we find that once the Fermi energy and $\nu$ are fixed, the magnetic field 
and the maximum density are restricted. 
The equation of state (\ref{eose}) then reduces to
\be
P=\frac{m_ec^{2}}{2K\Theta}\left(\rho \sqrt{K^2+\rho^2}-K^2\ln(\frac{\rho+\sqrt{K^2+\rho^2}}{K})\right)\label{1state}
\ee
for $\nu=0$.

For the two level system (when $\nu=0,1$)
\be 
n_{e}=\frac{\gamma}{2\pi^{2}\lambda^{3}}\sum_{\nu=0}^{1}\left(\left(2-\delta_{0,\nu}\right)\sqrt{x_{F}^{2}-2\gamma \nu}\right)=\frac{\gamma}{2\pi^{2}\lambda^{3}}\left(x_{F}+2\sqrt{x_{F}^{2}-2\gamma}\right)
\ee
and the pressure then reduces to
\be P=\frac{\gamma m_ec^{2}}{(2\pi)^{2}\lambda^{3}}\left(x_{F}\epsilon_{F}+2\epsilon_{F}\sqrt{x_{F}^{2}-2\gamma}-\ln(x_{F}+\epsilon_{F})-2(1+2\gamma)\ln(\frac{\sqrt{x_{F}^{2}-2\gamma}+\epsilon_{F}}{\sqrt{1+2\gamma}})\right)\label{2state}.\ee
The density can be written in terms of $x_F$ as
\be 
\rho=K\left(x_{F}+2\sqrt{x_{F}^{2}-2\gamma}\right).
\ee
Hence,
\be 
x_{F}=\frac{-\rho+2\sqrt{6K^{2}\gamma+\rho^{2}}}{3K}
\label{xf}
\eec
which is always positive definite. The other solution is neglected as infeasible solution as density cannot be negative for real particles.

\begin{figure}[]
\includegraphics[width=0.60\textwidth,angle=-90]{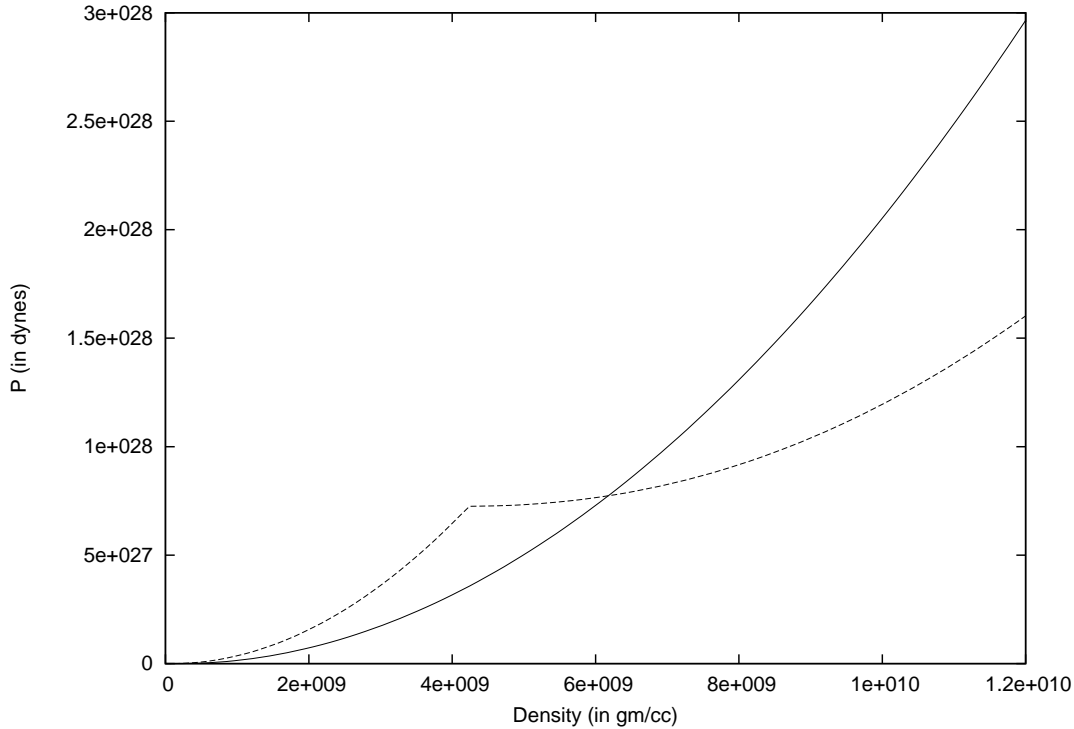}
\caption{Equations of state for one level (solid line) and two level (dashed line) systems as represented by equations (\ref{1state})
and (\ref{2state}) for $\epsilon_F=20$. The kink in dashed curve at $\rho_t$ is found from equation (\ref{ptne}).}
\label{2leveld}
\end{figure}

Substituting this $x_F$ in equation (\ref{2state}) we obtain
\begin{eqnarray}
\nonumber
P=\frac{m_e c^2 }{6 \Theta } \left(\epsilon_F  \left(6 K x_1-\rho +2 \sqrt{6 K^2 \gamma +\rho ^2}\right)-6 (K+2 K \gamma ) \ln\left(\frac{x_1+\epsilon_F }{\sqrt{1+2 \gamma }}\right)\right . \\
\left . -3 K \ln\left(\epsilon_F -\frac{\rho -2 \sqrt{6 K^2 \gamma +\rho ^2}}{3 K}\right)\right)\label{2pressure},
\end{eqnarray}
where \be x_1=\sqrt{-2\gamma+\frac{\left(\rho-2\sqrt{6K^{2}\gamma+\rho^{2}}\right)^{2}}{9K^{2}}}\ee
and 
\be \epsilon_{F}=\sqrt{1+\frac{\left(\rho-2\sqrt{6K^{2}\gamma+\rho^{2}}\right)^{2}}{9K^{2}}}.\ee
We must be careful that the two level system is only valid above a density given by equation (\ref{pt12e}). The net equation of state is represented by two functions given by:
\begin{eqnarray}
 &&
 {\rm Equation~~(16) }~~{\rm for}~~\rho < \rho_t~~(\nu=0) \\
\nonumber&&
 {\rm Equation~~(21)}~~~{\rm for}~~\rho > \rho_t~~(\nu=1). 
\label{2levele}
\end{eqnarray}
where $\rho_t$ is the density at which transition occurs between the Landau levels.
A diagrammatic representation of the equations (\ref{1state}) and (\ref{2state}) is given in Fig \ref{2leveld}.

Here the values of $K$ will depend on the limiting magnetic field taken. As for one level system the 
limiting magnetic field is much higher than that of two level system, the values of $K$ will be in 
general different between one level and two level systems. Hence, the steepness of the equation of state
representing ground Landau level will be different between the systems restricted to one level and that 
restricted to higher levels.
The higher level systems can be computed in a similar algebraic way. It must be noted that the value of 
$\gamma$ obtained from equation (\ref{fille}) corresponds to the lower limit for the given value of $\epsilon_F$.
For example, for say maximum Fermi energy $E_{Fmax}=20m_ec^2$, if $\gamma=199.5$ it corresponds to one level system, 
but for any other values greater than $199.5$ will not alter the condition for the same level system.

\subsection{Discussion of equation of states}
The electrons in the system become relativistic when the factor $\epsilon_F$ becomes more than one. 
For relatively low magnetic fields, the spacing between Landau levels becomes small and $\nu_{max}$ 
becomes large. Thus relativistic electrons can freely jump between the states as said earlier. 
In this case the levels can be treated as continuum and the summation over states can be replaced by 
an integral and the resulting equation gives back the normal non-magnetic relativistic equation of 
state $P=K'\rho^\Gamma$ with  $\Gamma=4/3$, just the case which was analysed by Chandrasekhar \cite{chandra}. Note that
$\Gamma=4/3$ corresponds to the equation of state of relativistic but non-magnetized electrons giving rise to the 
Chandrasekhar mass limit, 
which we recover in a particular range of density even in the magnetized system.
In just the opposite case, which we are interested in, we consider just one/two Landau level(s), i.e. 
the ground/first excited state, is filled in.  From equation (\ref{fille}) we fix $\nu_{max}$ to a 
value less than $1$ ($2$) such that only the ground level (and first level) is (are) occupied. The electrons in the 
ground level would all be 
in the spin $\downarrow$ state. From equation (\ref{fille}) we can see that once we fix $\nu_{max}$ and $x_F$, 
then the minimum magnetic field is automatically fixed. 

If there are finite Landau levels,  then the ground level would be filled first. If the density is increased 
further, then the ground state will saturate (it cannot contain more electrons) and there will be a phase 
transition like phenomena, in which electrons would start filling the first excited Landau level. This process will 
continue to higher Landau level transitions. During these phase transition like phenomena, the $E-\rho$ 
relation shows a sharp discontinuity. Thus $\frac{dE}{d \rho}=\infty$. An analytic condition can be found out 
for the critical densities at which this transition occurs by differentiating equation (\ref{energye}) with 
respect to $\rho$. In $\frac{dE}{d\rho}$ for two states, we note there is only one unique 
denominator which is not positive definite, hence can be set to zero which makes $\frac{dE}{d\rho}=\infty$. 
Thus this gives us the condition for the critical density. Solving the equation
\be 2\gamma =x_F^2\ee and equation (\ref{xf}) simultaneously
we obtain \begin{equation} \rho_t=\sqrt{2\gamma}K. \label{rhoce}\end{equation}
This gives the critical density below which the electrons cannot occupy the first Landau level ($\nu=1$) and 
would all be in the ground level ($\nu=0$) and hence it would 
be inappropriate to use the two level equation of state. This result can also be estimated by using 
$\sqrt{2\gamma}\le x_F$ which gives on combining with $x_F=\frac{\rho}{K}$ the same formula as equation 
(\ref{rhoce}). Similarly, the transition from first level to second level would then be given by the equation
\be \sqrt{2\times 2 \gamma} \le \frac{-\rho+2\sqrt{6 K^2\gamma+\rho^2}}{3 K}\label{pt12e}.\ee
The critical density for the transition from the first to second levels is given by
\be \rho_t = 2(1+\sqrt{2})\sqrt{\gamma}K.\ee  The higher phase change also occurs under the same principle.
Therefore we can write the formula of critical density for the transition from $(\nu-1)$-th level to 
$\nu$-th level for $1 \le \nu \le 2$ as
\be \rho_t(\nu)=(\sqrt{2\nu}+2\sqrt{2(\nu-1)})\sqrt{\gamma}K\label{ptne},\ee
where it must be remembered that $K$ is a function of $\gamma$.

\section{Determining mass and radius of the white dwarf analytically}\label{Ana}
For a magnetized star in hydrostatic equilibrium, we require to solve the condition for equilibrium given by 
\cite{shapiro}
\be\frac{1}{r^{2}}\frac{d}{dr}(\frac{r^{2}}{\rho}\frac{dP}{dr})=-4\pi G\rho(r), \label{hydroeqbme}\ee
where $P$ is the electron degeneracy pressure, varying with the radial coordinate of the star $r$,
playing the role to balance the gravitational force,
$G$ the Newton's gravitation constant.
Now we choose the equation of state as a polytropic relation 
\be P=K' \rho^ \Gamma \label{pe}\eec
where $K'$ is a dimensional constant, \be \rho=\rho_c \theta ^n \label{de} \eec and \be r=a\xi \label{re}\eec 
where $\rho_c$ is the central density of the white dwarf and $a$ is a constant defined by
\be  a=\left[\frac{(n+1)K^{'}}{4\pi G}\right]^{\frac{1}{2}}\rho_{c}^{\frac{(1-n)}{2n}}
\label{ae}.\ee  
With the use of equations (\ref{pe}), (\ref{de}), (\ref{re}) and (\ref{ae}), equation (\ref{hydroeqbme}) reduces to famous Lane-Emden equation 
\be\frac{1}{\xi^{2}}\frac{d}{d\xi}(\xi^{2}\frac{d\theta}{d\xi})=-\theta^{n}\label{lee}\eec 
where $\Gamma=1+\frac{1}{n}$. The Lane-Emden equation can be solved for a given $n$ using boundary conditions 
\be \theta(\xi=0)=1\ee and \be \left(\frac{d\theta}{d\xi}\right)_{\xi=0}=0.\ee 
For $n<5$, the value of $\theta$ falls to zero at a finite $\xi$ say $\xi_0$, which basically defines the surface of the 
star where pressure goes to zero. The physical radius is given by \be R=a \xi_0 \label{Re}.\ee 
Here we notice that $n\geq -1$ so that $a$ (see equation (\ref{ae})) must be real and so does  $R$.

Since the equation of state in presence of magnetic field cannot be written in a simple polytropic form as 
in equation (\ref{pe}), we fit equation (\ref{eose}) for various density ranges. The actual equation of state 
is  reconstructed by using multiple fits in various density range. The values for $K'$ and $\Gamma$ 
in various ranges of density, which are found from the fitting function, also carry information about the 
magnetic field of the system. The idea behind the fit is to be able to solve the Lame-Emden equation 
piecewise ranges of density and to obtain an idea of the mass-radius relation from it easily. The mass of 
the star is then given by 
\be M=\int4\pi\rho r^{2}dr = 4\pi\rho_c a^3\int^{\xi_0}_0 \xi^2 \theta^n d\xi.\label{masxi}\ee

\begin{figure}[]
\includegraphics[width=0.60\textwidth,angle=-90]{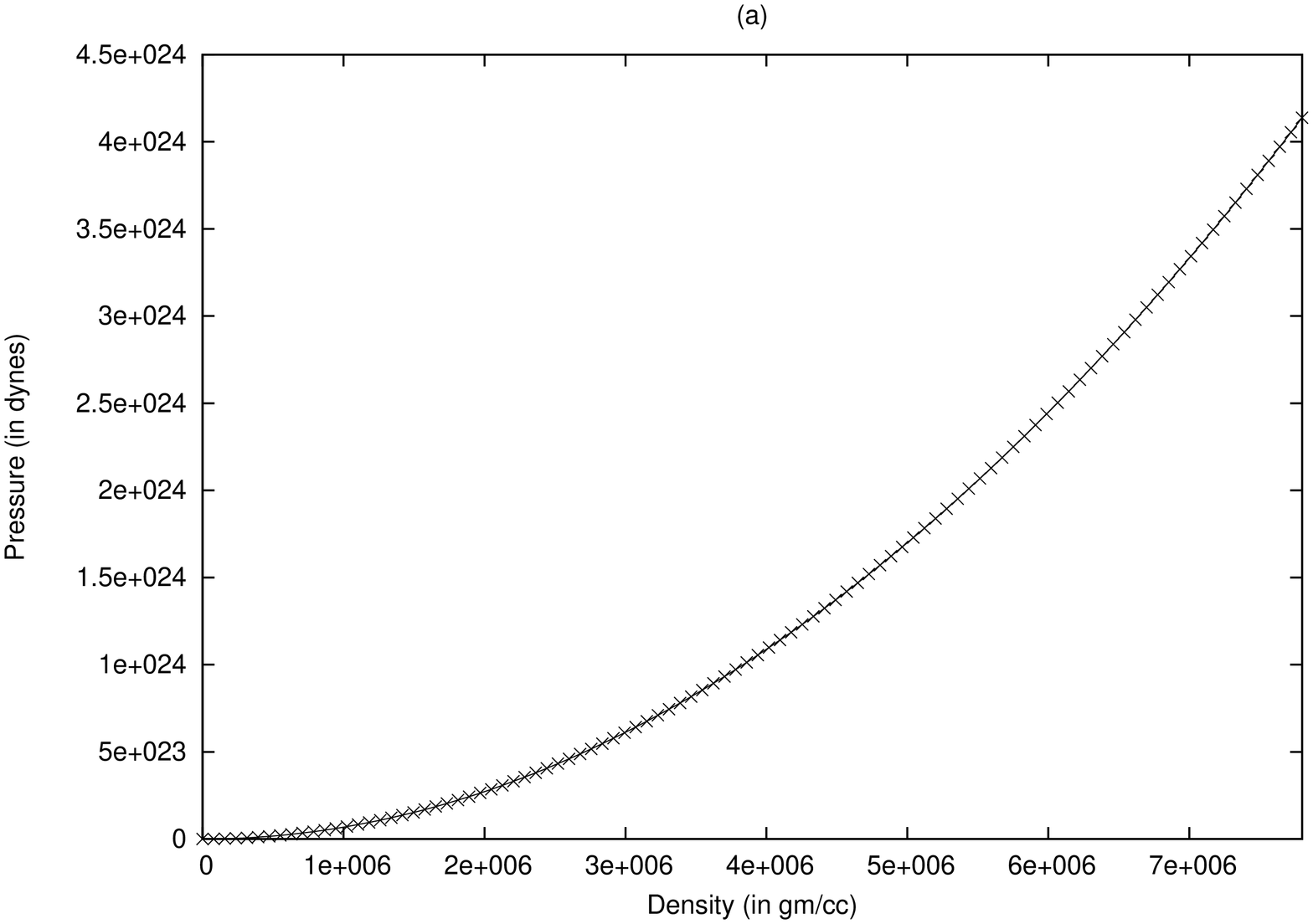}
\includegraphics[ width=0.60\textwidth,angle=-90]{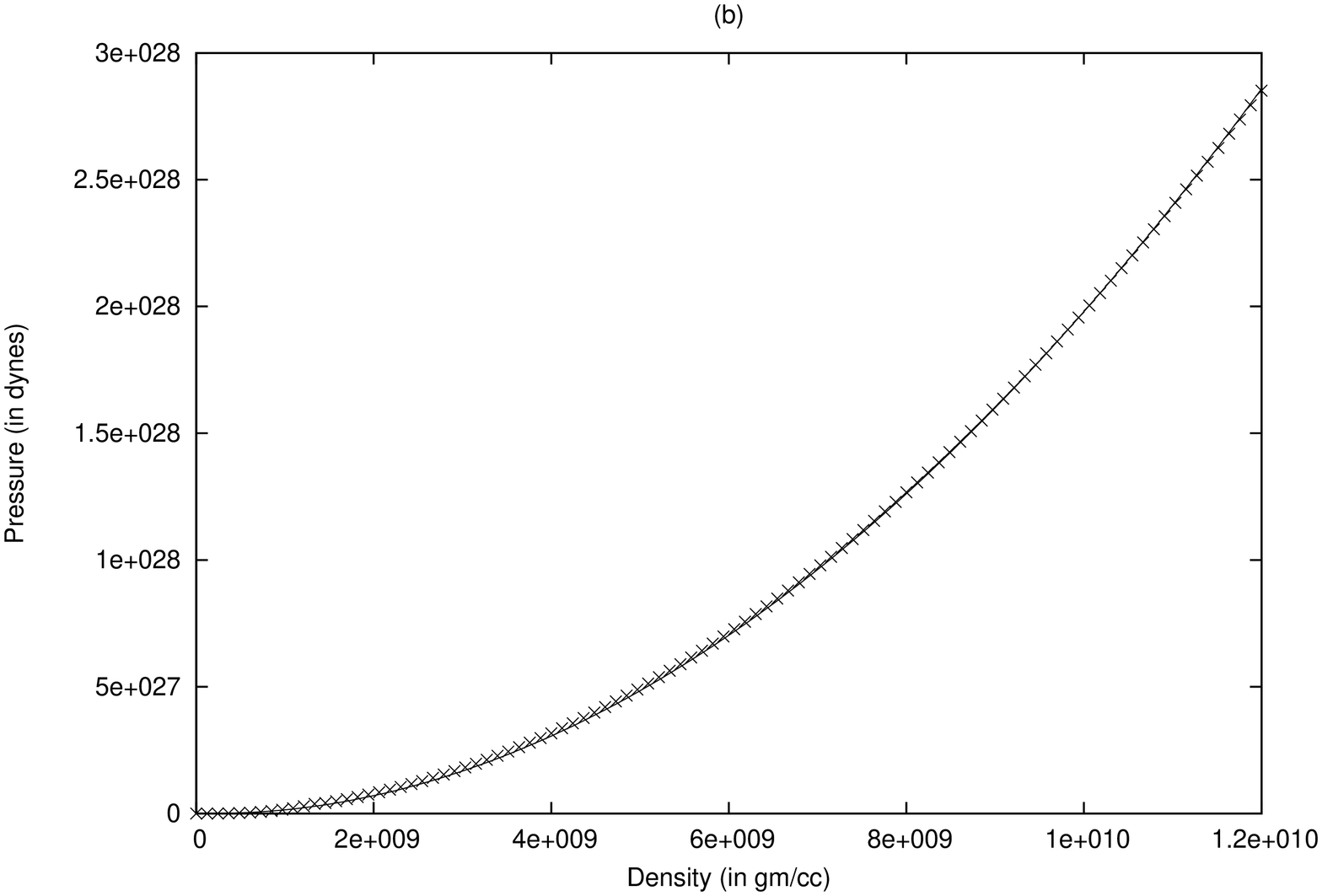} 

\caption{Equations of state for one Landau level system, when (a) $\epsilon_F=2$, (b) $\epsilon_F=20$.
Solid and broken lines respectively indicate original and fitting results, both overlap each other perfectly. }
\label{EoS1}
\end{figure}

Now we consider only one level system explicitly, hence fit the one level equation of state and obtain 
different values of $K'$ and $\Gamma$. 
We see from Fig. \ref{2leveld} that the equation of state representing the one level system for $\epsilon_F=20$
is given by 

\be P=2\times 10^8\left(\rho\sqrt{34\times 10^{16} +\rho^2}-34 \times 10^{16} \ln\left((\rho+\sqrt{34\times 10^{16}+\rho^2})~16\times10^{-10}\right)\right) \eec

which is obtained from equation (\ref{1state}). 
This curve is fitted with two functions
\begin{eqnarray}
\nonumber &&
P(\rho)=  6.98\times 10^{-1} \rho^{2.92}~~~{\rm for}~~~\rho  \leq\: 1.4\times 10^9,\\
\nonumber&&
P(\rho)=  1.98\times 10^{8} \rho^2~~~~~~~{\rm for}~~~\rho   \geq\: 1.4\times 10^9, 
\label{2levelfit}
\end{eqnarray}
shown by Fig. \ref{EoS1}(b).
Here we consider a uniform magnetic field. In very low densities where we consider the pressure 
goes to zero, the equation of state defined above cannot be used as it becomes non-linear in nature. 
We have extended the fitting curve until this region arises.
However, this region is a very small part of the total density region, hence would not affect our results 
effectively. Similarly, the fitting of the one level equation of state for $\epsilon_F=2$
is given by Fig. \ref{EoS1}(a).
Then we integrate equation (\ref{masxi}) in two ranges of density to obtain
\be M=4\pi\rho_c \left( a(n_1)^3\int^{\xi_i}_0 \xi^2 \theta^n d\xi  + a(n_2)^3(\int^{\xi_0}_{0} \xi^2 \theta^n d\xi - \int^{\xi_i}_0 \xi^2 \theta^n d\xi )\right)\eec
which gives
\be M=4 \pi \rho_c \left({a(n_1)}^3|{\xi_i}^2\theta^{'}(\xi_i)|+{a(n_2)}^3(|{\xi_0}^2\theta^{'}(\xi_0)|-|{\xi_i}^2\theta^{'}(\xi_i)|)\right) \label{mie}\eec
where $\xi_i$ is the radius corresponding to the intermediate density where the $\Gamma$ changes its value 
drastically in the equation of state. We then substitute this intermediate density ($\rho(\xi_i)$) in equation (\ref{de}) with a 
particular value of central density and corresponding polytropic exponent and find the corresponding $\theta$. 
From the solution to Lane-Emden equation we find the corresponding values of $\xi$ and ${\xi}^2|\theta^{'}(\xi)|$.
The radius in this case will be
\be R=a(n_1) \xi_i +a(n_2) ( \xi_0- \xi_i) \label{rie}\eec
where the values of $a(n_1)$ and $ a(n_2)$ are obtained from equation (\ref{ae}) by substituting the 
corresponding values of $n$ and $K^{'}$.

\section{Mass-Radius relation}

\begin{figure}[]
\includegraphics[width=0.60\textwidth,angle=-90]{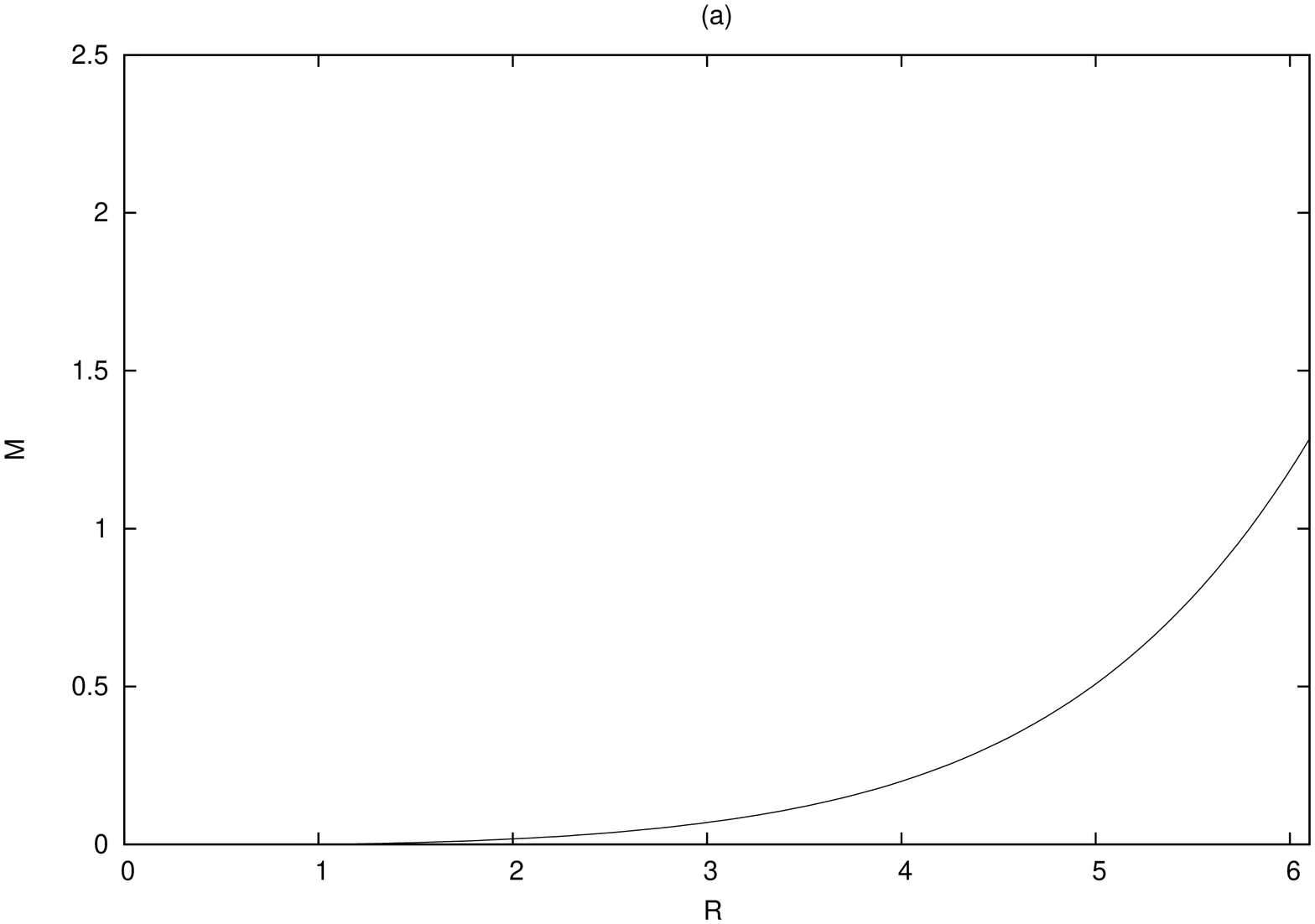} 

\includegraphics[width=0.60\textwidth,angle=-90]{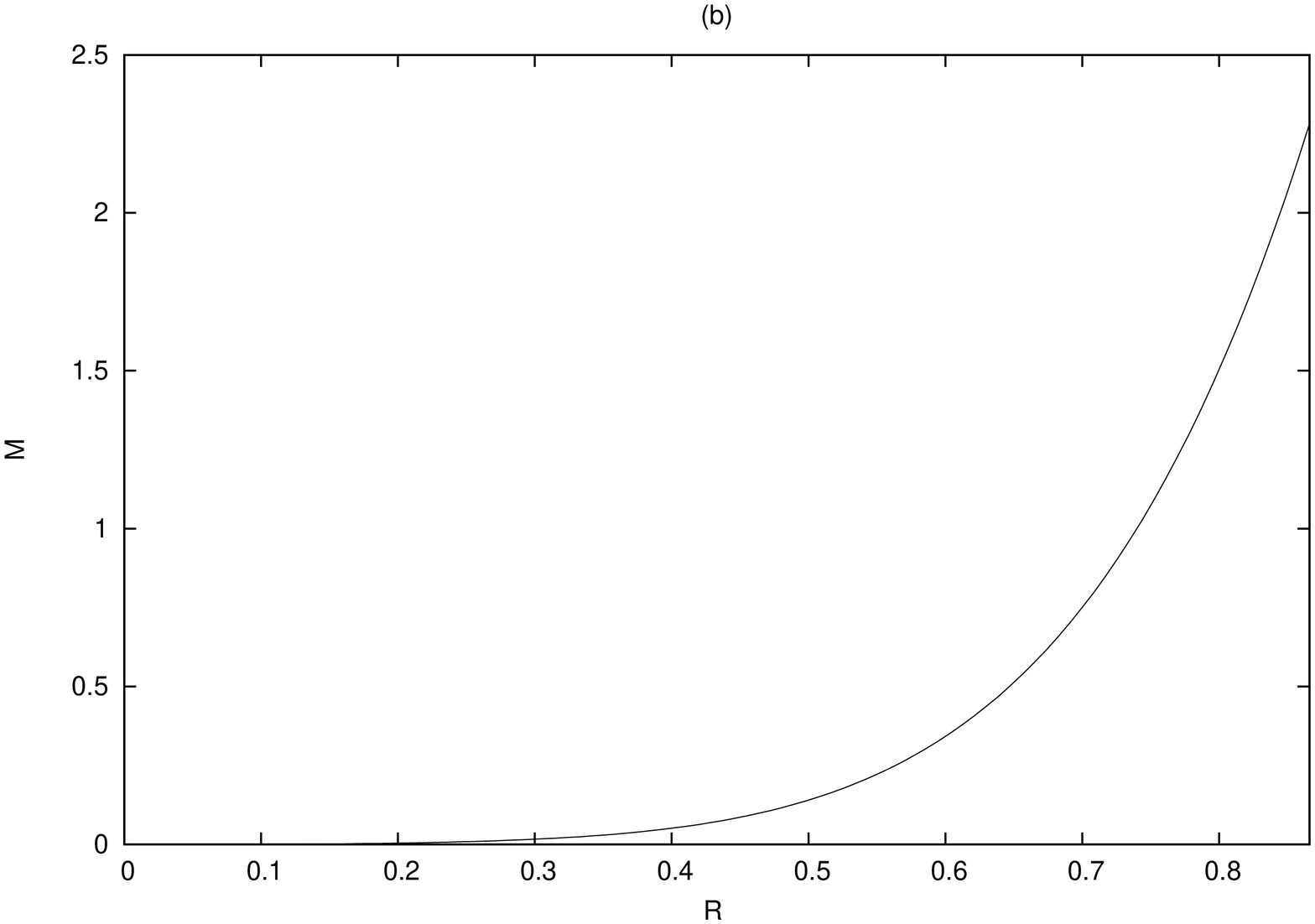} 
\caption{Mass-Radius relation for magnetic white dwarfs for one Landau level systems, when (a) 
$\epsilon_F=2$, (b) $\epsilon_F=20$. $M$ is expressed in the units of $M_\odot$ and $R$ in the units of $10^8~cm$. 
The maximum 
mass is determined from the maximum value of $\rho_c$ for one level system as determined from equation (\ref{fille}).
For $\epsilon_F=2 $, the corresponding $\rho_c$ is $7.4 \times 10^6 gm/cc $ and for $\epsilon_F=20$, 
the corresponding $\rho_c$ is $1.2 \times 10^{10} gm/cc$.}
\label{MR}
\end{figure}

We now discuss the mass-radius relation for the highly magnetized star. 
We again consider to study the extreme case having only one Landau level. For the higher level occupancy, 
as shown by the equation of state in Fig. \ref{2leveld} by the dashed curve for a two level system, 
the magnetic effect will be 
softened and we will discuss the results qualitatively. 
From equation (\ref{mie}) we obtain mass as a function of central density and equation 
(\ref{rie}) gives the radius of the star as a function of central density, for one level systems. 
The values of coefficients 
are calculated from the values of $K'$ and $\Gamma$ from the fitting curves in various zones and 
$\gamma$ is determined from the Fermi energy.
We can eliminate the central density from equations (\ref{mie}) and (\ref{rie}) and obtain mass as a 
function of radius. In order to express the results in convenient units we write the mass in the units of 
solar mass and radius in the units of $10^8 ~cm$.
The values of $K'$ and $\Gamma$ obtained from the fitting are used in equations (\ref{mie}) and (\ref{rie}) 
which gives for $\epsilon_F$ =20 
\be M/\Msun = \: 3.81291\times 10^{-11} \rho_c \, + 2.86571 \times 10^{-23} \rho_c^{2.38} \ee
and
\be R/10^8=\: 0.05839 \,+ 3.196\times 10^{-5} \rho_c ^{0.46}.\ee
%
Now eliminating $\rho_c$ from above equations we obtain the mass-radius relation.
Figure \ref{MR} shows the mass-radius relation for $\epsilon_F=2,20$.
The extreme right point of the curves (maximum radius) denotes the typical maximum central density 
of white dwarfs for the respective values of $\epsilon_F$. Thus the mass corresponding to that radius gives the maximum mass 
possible for the white dwarf.

We know from the non-magnetic case that a stiffer equation of state can balance gravity more effectively 
leading to a higher value of radius (non-relativistic case). The same argument can be applied here for 
$\Gamma=2.92$ (lower density and hence outer region of the star), owing to the fact that the stiffness arises from the effect of magnetic field on degeneracy 
pressure. Hence from the mass-radius relation, mass would turn out to be higher than the usual one at a 
particular radius. Here we must be careful that even though the resulting analytical form of equation apparently does not have any 
restriction on the value of the maximum mass or radius, the maximum mass is restricted by the typical values 
of central density for white dwarf \cite{shapiro} and the maximum Fermi energy of the star determined by equation 
(\ref{fille}). Figure \ref{MR} however shows that
the magnetic white dwarfs have typically higher mass and higher radius. In general 
for typical densities of white dwarf, the mass of the white dwarf can exceed the Chandrasekhar mass limit. 
For $\epsilon_F=2$, however, we see from the mass-radius relation that the typical radius turns out to be 
higher than that of $\epsilon_F=20$, but the Chandrasekhar mass limit is not exceeded.

In the equation of state for one level, $\Gamma$ decreases from $2.92$ in the low density region to the 
value $2$ in the high density region. 
Quantitatively for a two level system, we can see from the equation of state that after the transition from 
the ground level to the first level, the slope of the curve changes drastically to very low values and then 
rises again. At this transition point, the pressure is nearly independent of density. This is physically
improbable and the corresponding mass-radius relation would give an unstable branch. 
If a star starts out in this density range, it cannot form a compact star and will go into a runaway 
process.

\section{Discussion and summary}
In this work we have calculated the equation of state of a degenerate electron gas in high magnetic field at 
zero temperature analytically. For simplicity, and keeping it analytically tractable, 
we have discussed the extreme cases when there are only one 
and two Landau levels for the system and studied the mass-radius relation for the case of one Landau level.
Our pure analytical approach helps in understanding the underlying physics in great details.
In the forthcoming work, we will show that the detailed numerical unrestricted solutions indeed match with 
our analytical results \cite{dm}.

To make the problem theoretically tractable easily, we have chosen constant magnetic field throughout. However, this
does not matter for the present purpose due to the following reason.
Although the matter density and the magnetic field both vary from center (highest) to surface (lowest),
by the time the density falls to about half the value of
the central density, the mass generally crosses the Chandrasekhar mass
limit. Thus, although we have considered a constant magnetic field, the field strength plays its major role and brings in
new results in the high density regions of the star only. Hence it appears equivalent to the central magnetic field
of the star. Now following previous work \cite{prl,bani},  one can adopt an inhomogeneous
magnetic field profile in any compact star which is nearly constant throughout
most of the star and then gradually falls off close to the surface (see Figure
 5(b) in \cite{bani}). Thus the choice of an inhomogeneous magnetic field would not affect
our main finding. A detailed description for the same will be given in a follow up paper based on detailed numerical analyses 
\cite{dm}. 

Note that the size of a white dwarf is very large so that the effect of general relativity is insignificant there.
In addition, the central density of the star is almost an order of magnitude larger than the energy density arised 
due to the corresponding magnetic field. As the effect of magnetic field arises in the high density regime only when the 
matter density exceeds the energy density arised due to the magnetic field, even the magnetic field does not contribute to 
the gravitational field leaving the system Newtonian.

It is found that the ground level of the system is occupied 
first and then the higher levels start getting filled in. This transition appears as a kink 
in the pressure-density plot, what we have addressed analytically. For one level system that we have studied 
explicitly for the mass-radius relation (our numerical solutions addressed in a separate paper \cite{dm} would show the
mass-radius relations for multi level systems), it is known to have no kink, for two level systems one kink, and 
so on for higher level systems, what we have studied analytically. We have derived a general expression for the positions of the kinks in the
equation of state in 
terms of the magnetic field and other constants and the relation among them. It has been found that 
they bear a simple ratio, thus if we know exactly the value of density for a kink, then we can find out 
the density for other kinks. One of the cases considered here is for the Fermi energy $20m_ec^2$ and 
the system is of a one level.
The minimum magnetic field required for this system is found to be $8.7 \times 10^{15}~G$. The corresponding 
mass of the resulting white dwarf has been found to be larger than that predicted by Chandrasekhar and is 
about $2.3\Msun$ when the radius is $8\times 10^7~cm$. We have also analysed analytically how is the mass-radius 
relation dependent on the Fermi energy and the magnetic field. 

We end by addressing the possible reason for not observing such a high field yet in a white dwarf. This 
could be due to the magnetic screening effects on the surface of the star. If the white dwarf is an accreting one, 
then the current in accreting plasma depositing on the surface of the white dwarf could create an induced magnetic moment 
of sign opposite to that of the original magnetic dipole, thus reducing the surface magnetic field of the white dwarf 
unaffecting the central field. In addition, the surface field could be several orders of magnitude smaller 
than the field in the central region which brings in the main results. Hence by estimating the surface field 
one should not interpret the rest.
\\ \\

{\bf Acknowledgments}: This work was partly supported by an ISRO grant ISRO/RES/2/367/10-11 
and a KVPY grant of DST. We thank Upasana Das for discussion and cross checking several calculations.

\end{document}